\documentclass[preprint,endfloats,prl,showpacs]{revtex4}

\usepackage{amsmath,graphicx,natbib}

\newcommand{\CeCoIn}{CeCoIn$_5$}
\newcommand{\CeCoSn}{CeCoIn$_{5-x}$Sn$_x$}

\begin{document}

%\wideabs{

\title{Superconductivity in CeCoIn$_{5-x}$Sn$_x$:  Veil Over an Ordered State or Novel Quantum Critical Point?}

\author{E. D. Bauer,  C. Capan, F. Ronning, R. Movshovich, J. D. Thompson, J. L. Sarrao}
\affiliation{Los Alamos National Laboratory, Los Alamos, New Mexico 
87545, USA}

%\draft

\date{\today}

%\maketitle

\begin{abstract}

Measurements of specific heat and electrical resistivity in magnetic fields up to 9 T along [001] and temperatures down to 50 mK of Sn-substituted CeCoIn$_5$ are reported.    
The maximal $-ln(T)$ divergence of the specific heat at the upper critical field $H_{c2}$ down to the lowest temperature characteristic of non-Fermi liquid systems at the quantum critical point (QCP), the universal scaling  of the Sommerfeld coefficient, and agreement of the data with spin-fluctuation theory, provide strong evidence for quantum criticality at $H_{c2}$ for all $x\leq 0.12$ in \CeCoSn.  These results indicate the ``accidental" coincidence of the QCP located near $H_{c2}$ in  pure \CeCoIn, in actuality, constitute a novel   quantum critical point associated with unconventional superconductivity.  

\end{abstract}

%}

\maketitle

% 74.70.Tx Heavy-fermion superconductors
% 65.40.-b Thermal properties of crystalline solids
% 71.27.+a Strongly correlated electron systems; heavy fermions
% 75.30.Mb Valence fluctuation, Kondo lattice, and heavy-fermion phenomena 

The emergence of  exotic types of order  at the boundary separating an ordered phase from a disordered one at zero temperature, or quantum critical point (QCP), is the current subject of intense experimental and theoretical research.  
Attention has focussed on antiferromagnetic quantum critical points in $f$-electron heavy fermion materials, leading to the discovery of superconductivity near the suppression of the N{\'{e}}el temperature in such systems as  CeIn$_3$ and CePd$_2$Si$_2$ \cite{Mathur98}.  
More recently,  novel ground states have been found in proximity to a variety of QCPs associated with  ``hidden" order \cite{hiddenorder2} (e.g., URu$_2$Si$_2$), quadrupolar order \cite{Sakakibara04} (e.g., PrFe$_4$P$_{12}$), metamagnetism \cite{Grigera01} (e.g., Sr$_3$Ru$_2$O$_7$), or helimagnetism \cite{Pfleiderer04} (e.g., MnSi).  In this Letter, we investigate another type of QCP, namely quantum criticality associated with suppression of unconventional superconductivity in \CeCoIn.

Various control parameters such as pressure, composition, and magnetic field have been used to tune systems through their respective QCPs.  At this point, the spectrum of abundant low-energy quantum fluctuations leads to a striking departure from typical metallic behavior characterizing a Fermi liquid [Sommerfeld coefficient $C/T \sim const.$, magnetic susceptibility $\chi \sim const.$, and electrical resistivity $\rho(T) = \rho_0 + AT^2$].  Instead,  in the vicinity of the QCP the system  exhibits non-Fermi liquid (NFL) behavior, i.e., $C/T \sim -lnT$, $\chi \sim T^{-n}$ ($n  <1$), and  $\rho(T) = \rho_0 + AT^n$ ($n<2$) \cite{Stewart01}.

Attention has been lavished on the quasi-2D heavy fermion superconductor \CeCoIn{} due to its unusual normal and superconducting states \cite{Petrovic01}.  Superconductivity in this material observed at $T_c=2.3$ K  is unconventional, as evidenced by the power-law behaviors of its thermal conductivity, specific heat, and spin-lattice relaxation  \cite{Movshovich01,Kohori01}.  
Furthermore, magnetic-field dependent thermal conductivity experiments \cite{Izawa01} are consistent with  $d$-wave  superconductivity.  
The first-order nature of the superconducting transition in magnetic fields and a second anomaly observed close to $H_{c2}$ below 1 K  make \CeCoIn{} an excellent candidate for a Fulde-Ferrell-Larkin-Ovchinnikov (FFLO) state \cite{Bianchi03b,Radovan03,FFLO}.  The normal state of \CeCoIn{} is equally unusual, characterized by a NFL $C/T \sim - lnT$ and a $T-$linear electrical resistivity \cite{Petrovic01}, consistent with proximity to an antiferromagnetic (AFM) QCP \cite{Millis93}.  Further, measurements \cite{Bianchi03} in magnetic fields above $H_{c2}=4.95$ T ($H||c$) reveal an evolution from NFL to FL behavior and a universal scaling of the Sommerfeld coefficient, leading to the conclusion that long-range AFM order was narrowly avoided at a quantum critical point $H_{QCP}=5$ T.  A  similar evolution exists in \CeCoIn{} when the magnetic field is applied in the $ab$-plane where $H_{c2}=12$ T \cite{Ronning04}.
%The unexpected coincidental suppression of both AFM and superconducting order in \CeCoIn{} is a rather intriguing and yet unresolved issue in heavy fermion physics.

The \CeCoSn{} system is ideally suited to address the issue of the coincidental nature of the suppression of superconductivity and the quantum critical point as superconductivity is rapidly suppressed at a rate $dT_c/dx = -0.6 $K$/$at.\% Sn while the electronic structure remains essentially unchanged \cite{Bauer04d}.  
A shift of the QCP away from $H_{c2}$ in \CeCoSn{} should be readily observable:  if $H_{QCP}$ moves in the superconducting region, robust Fermi-liquid behavior will occur either above $T_c$ or $H_{c2}$; in contrast, if superconductivity is suppressed more quickly than the QCP, long-range magnetic order will be revealed.  
In this Letter,   we describe in detail a remarkable and completely unexpected result based upon $C(H,T)$ and $\rho(H,T)$ measurements:  {\it{while there is no sign of long-range magnetic order, we cannot distinguish experimentally between the quantum critical point at $H_{QCP}$ and the destruction of superconductivity at $H_{c2}$ in}} \CeCoSn{} {\it{for all Sn concentrations investigated ($x\leq0.12$)}}.  Thus, the occurrence of quantum criticality and suppression of superconductivity at  $H_{c2}$ in \CeCoIn{} is not ``accidental", but is a manifestation of the underlying physics.  We discuss various theoretical scenarios consistent with our results.
%Fig 1
\begin{figure}[htbp] 
\includegraphics[width=5in]{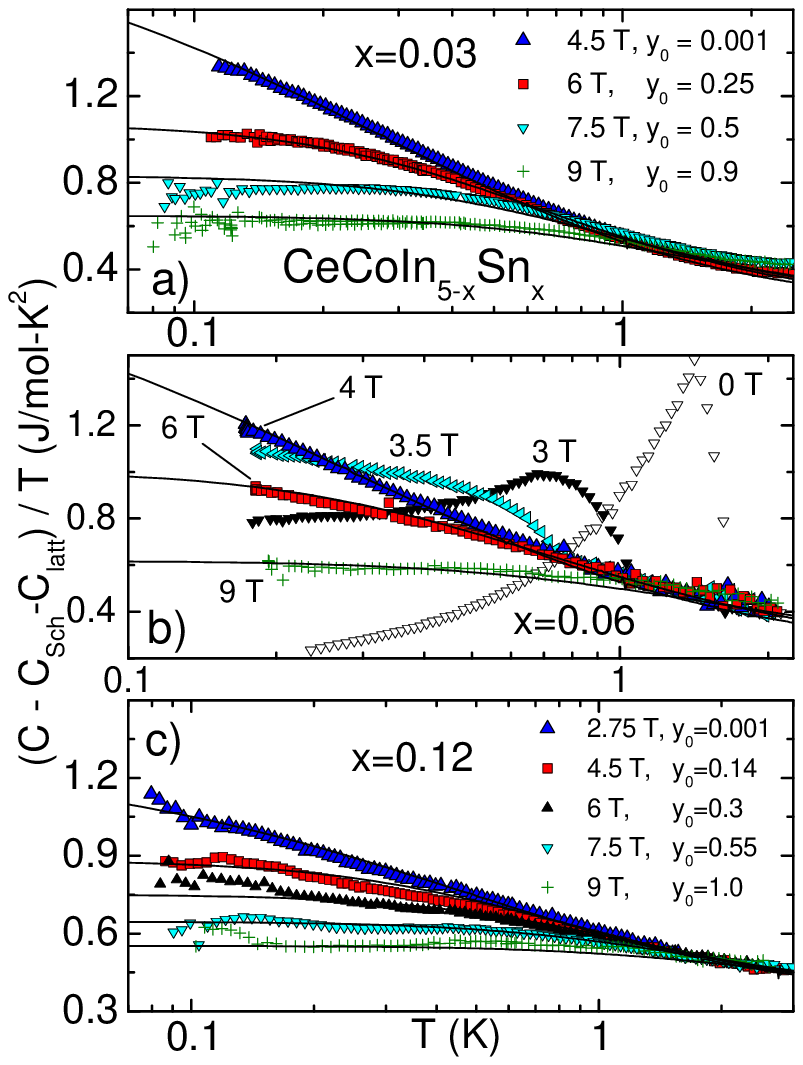}
\caption{(color online) Electronic contribution to the specific heat $C_{el}= C - C_{Sch} - C_{latt}$ divided by temperature $T$  of \CeCoSn{} in magnetic fields $H||[001]$  for a) $x=0.03$, b) $x=0.06$, and c) $x=0.12$.  The solid lines are fits of the  spin-fluctuation theory \cite{Moriya95} discussed in the text [$y_0$=0.001, 0.3, 1.0 for 4 T, 6 T, and 9 T, respectively in b)].}
\label{celot}
\end{figure}

Single crystals of CeCoIn$_{5-y}$Sn$_y$ ($0 \leq y \leq 0.4$) were grown in In flux in the ratio Ce:Co:In:Sn=1:1:20:$y$.  Microprobe analysis (MPA) reveals a Sn concentration $x$ $\sim 0.6y$; hereafter, the actual values ($x$) deduced from MPA rather than the nominal values ($y$) will be quoted.

We focus our attention on specific heat measurements in magnetic fields up to 9 T ($H||[001]$) and down to 50 mK of \CeCoSn{} for $x \leq0.12$.  The electronic contribution to the specific heat $C_{el}/T$ is shown in Fig. \ref{celot}, where the lattice contribution of nonmagnetic LaCoIn$_5$, $C_{latt}$, and a low-$T$ Schottky anomaly tail arising from the splitting of degenerate Co and In nuclei with $H$, $C_{Sch}$, have been subtracted from the data \cite{Movshovich01}.  For magnetic fields $H=H_{c2}$ [=4.5 T, 4 T, and 2.75 T for $x=0.03$, 0.06, and 0.12, respectively], the data exhibit a logarithmic divergence below 1 K down to the lowest measured temperature, characteristic of NFL systems in the vicinity of a QCP.  With increasing field, $C_{el}/T$ deviates from the $-ln(T)$ dependence at low temperature; a crossover region at $T_{cr}$ can be identified for these intermediate fields, while evidence for Fermi-liquid behavior ($C_{el}/T \sim const.$) is found only for $H\geq7.5$ T for all $x$.  As shown in Fig. \ref{celot}b, when superconductivity is suppressed,  the NFL behavior persists to the lowest temperature indicating that superconductivity develops out of a NFL ground state.  It is interesting to note that the superconducting transition is always second-order ($x>0$).  No evidence of magnetic order is observed in these $C(H,T)$ [or $\rho(H,T)$] measurements.  Taken together, the lack of magnetic order and the fact that the strongest divergence of $C_{el}/T$ is found at $H_{c2}$  implies that the QCP is closely related to the complete suppression of superconductivity.

\begin{figure}[htbp] 
\includegraphics[width=6in]{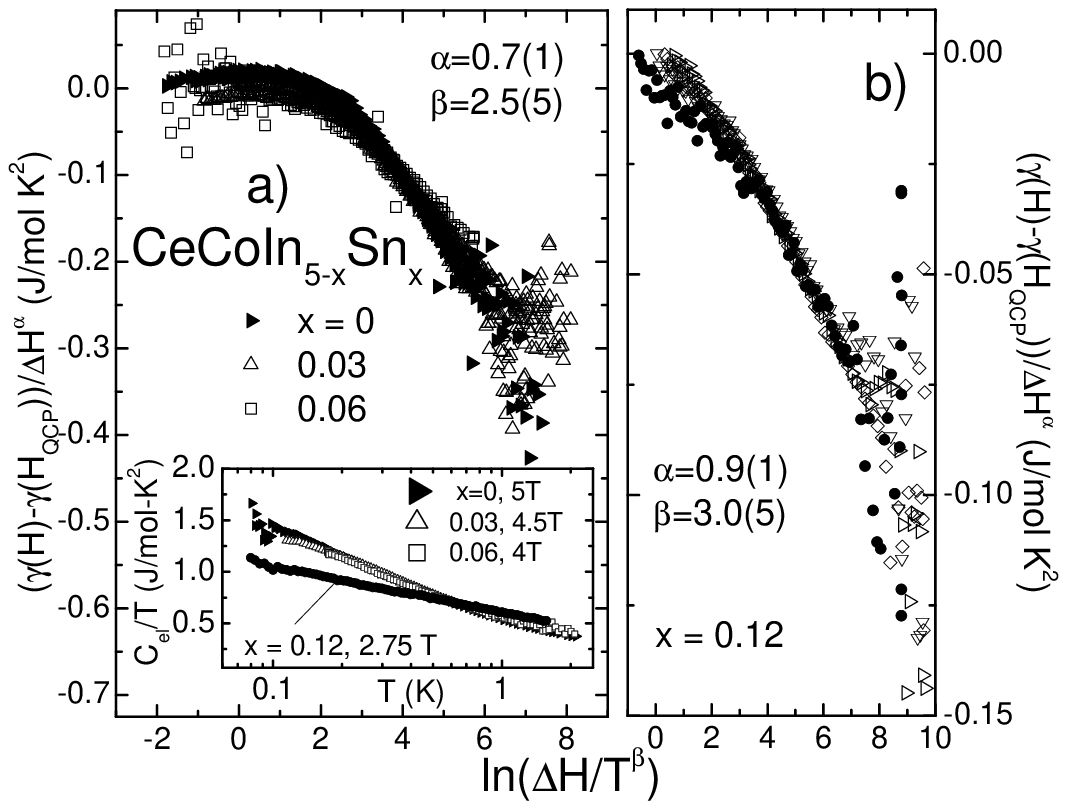}
\caption{Scaling analysis of the Sommerfeld coefficient $\gamma$ of \CeCoSn{} for a) $x=0$ \cite{Bianchi03}, 0.03, 0.06, and b) $x=0.12$.  Inset a) $C_{el}(T)/T$ at $H=H_{c2}$ for $x \leq 0.12$.}
\label{scaling}
\end{figure}

Further support of a QCP at $H_{c2}$ is provided by the  scaling analysis of the Sommerfeld coefficient as shown in Fig. \ref{scaling}.
The $C_{el}/T \equiv \gamma$ data in applied fields for $x=0,0.03,0.06$, plotted as $[\gamma(H) - \gamma(H_{QCP})]/(\Delta H)^{\alpha}$ vs $\Delta H/T^{\beta}$ with $\Delta H=(H - H_{QCP})$, can be collapsed onto a single curve choosing $H_{QCP}= H_{c2}$ (Fig. \ref{scaling}a) and critical exponents $\alpha=0.7(1)$ and $\beta=2.5(5)$.  While the data for $x=0.12$ could be included on this plot, a choice of critical exponents $\alpha=0.9(1)$ and  $\beta=3.0(5)$ better describe the data (Fig. \ref{scaling}b), possibly indicating the influence of disorder on the scaling.  Such scaling  has been observed in other NFL systems such as U$_{0.2}$Y$_{0.8}$Pd$_3$ \cite{Andraka91} and YbRh$_2$Si$_2$ \cite{Trovarelli00}, and is viewed as evidence for proximity to a QCP.  The inset of Fig. \ref{scaling}a shows the striking similarity of the Sommerfeld coefficient at criticality (i.e, $H_{c2}$) for $x=0, 0.03, 0.06$, while the $x=0.12$ sample exhibits a  $-ln(T)$ divergence with a smaller slope.
%Fig 2

\begin{figure}[htbp] 
\includegraphics[width=6in]{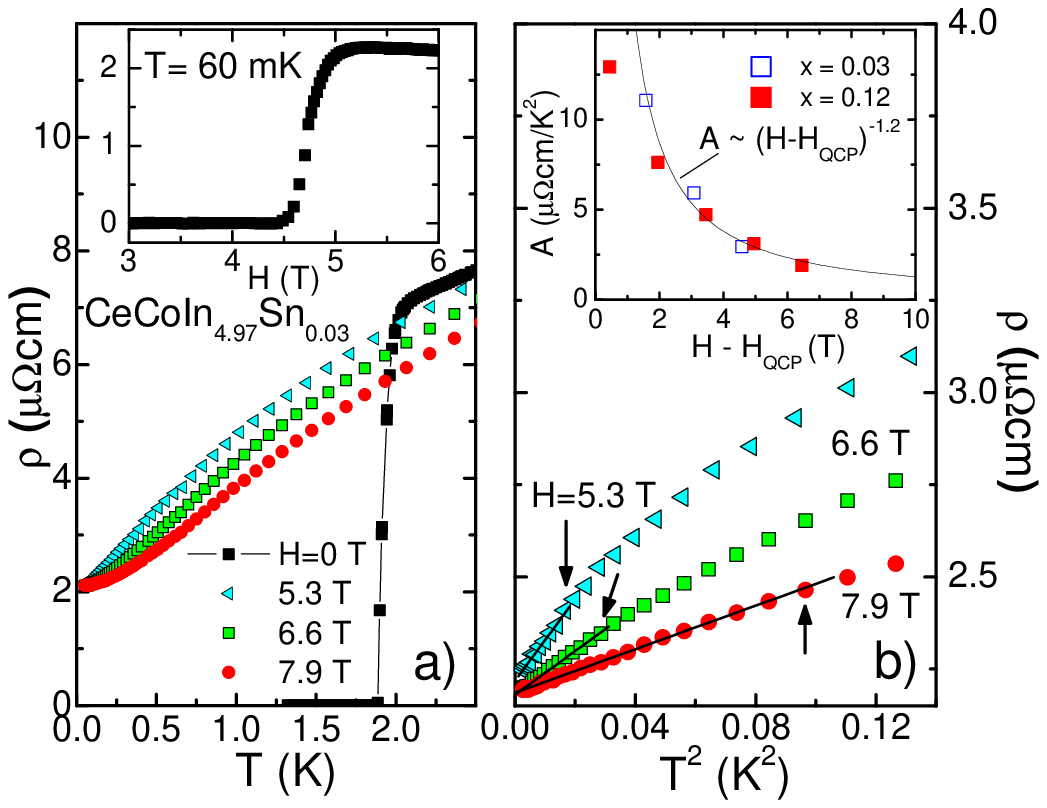}
\caption{(color online) a) Electrical resistivity $\rho(T)$ of \CeCoSn{} for $x=0.03$ for $H||[001]$. Inset:   $\rho(H)$ at $T=60$ mK. b) $\rho$ vs $T^2$ for data in a).  The solid lines are linear fits to the data. The arrows denote the maximum temperature $T_{FL}^{\rho}$ of the Fermi liquid behavior. Inset:  $A$ vs $H-H_{QCP}$ ($H_{QCP}=H_{c2}$) for $x=0.03$ and $x=0.12$.  The solid line is a power law fit to the $A(\Delta H)$ data for $x=0.03$.}
\label{mr}
\end{figure}

The electrical resistivity $\rho(T)$  for \CeCoSn{} for $x=0.03$ in  applied fields  is shown in Fig. \ref{mr}a.  At $H=5.3$ T, $\rho(T)$ follows a NFL power law $T$-dependence $\rho - \rho_0 = AT^n$ with $n=1.5(1)$ over nearly a decade in temperature from 0.05 K to 0.4 K.
% consistent with a divergent Sommerfeld coefficient close to this field at $H_{c2}=4.5$ T. 
  The $\rho(T)$ data can also be analyzed for $H\geq H_{c2}$ by a Fermi liquid  form $\rho(T) = \rho_0 + AT^2$ as displayed in Fig. \ref{mr}b, yielding a rapid decrease in   $A$  away from  $H_{c2}$ (inset of Fig. \ref{mr}b) (similar behavior is also found for $x=0.12$).   A power law fit to the data of the form $A(H) \sim 1/(H-H_{QCP})^{\alpha}$ gives $\alpha=1.2$ (1.1) for $x=0.03$ ($x=0.12$) with $H_{QCP}=H_{c2}$ similar to  \CeCoIn{} \cite{Paglione03}.  It is possible that the data closest to the QCP cannot accurately be described by this $T^2$ behavior, thus leading to a deviation from the divergent power law dependence of $A$.  
%Fig 3

To further analyze the $C(H,T)$ and $\rho(H,T)$ data and to gain information about the distance from the quantum critical point in applied field, we employ the spin-fluctuation theory of Moriya and Takimoto \cite{Moriya95}.  In this model, anomalous NFL $T$-dependences of $C(T)$ and $\rho(T)$ due to critical AFM spin fluctuations are calculated as a function of reduced temperature $T/T_0$, with a control parameter $y_0$ denoting the distance from the QCP (i.e., $y_0$=0 at the QCP) that provides a measure of the inverse correlation length.  Two additional parameters are needed for comparison to experiment.  The first parameter $T_0$, is related to the exchange energy by $T_0 =\mathcal{J}/(2\pi^2)$, which we take to be close to that of the N{\'e}el temperature $T_N \approx$ 4 K of the homologous compound CeRhIn$_5$ \cite{Hegger00}, and the second is the contribution to the electronic specific heat of non-critical fermions $\gamma_0$ of the order of the Sommerfeld coefficient at $T_c$.  The fits of $C_{el}/T$ data of \CeCoSn{} to the Moriya-Takimoto model are shown in Fig. \ref{celot}.  We emphasize that data collected at $H_{c2}$ are closest to the QCP (all $y_0 \leq 0.01$ describe these data well), with a smooth evolution away from the critical point in higher magnetic field.  These fits  support our assertion that quantum criticality occurs at $H_{c2}$ for all $x$ in \CeCoSn.   Identical parameters $T_0=0.4$ K and $\gamma_0=0.2$ J/mol K$^2$ are obtained for $x=0$ (not shown) \cite{Bianchi03}, 0.03, and 0.06;  slightly different parameters $T_0=0.7$ K and $\gamma_0=0.34$ J/mol K$^2$ are needed to characterize the $x=0.12$ sample.  The ``s"-shaped curvature of $\rho(T)$ of \CeCoSn{} below 2 K is reasonably well reproduced by the spin-fluctuation theory using identical parameters determined from $C(H,T)$ measurements (not shown) and any discrepancy between theory and experiment likely arises from disorder effects  not included in the model.

\begin{figure}[htbp] 
\includegraphics[width=5in]{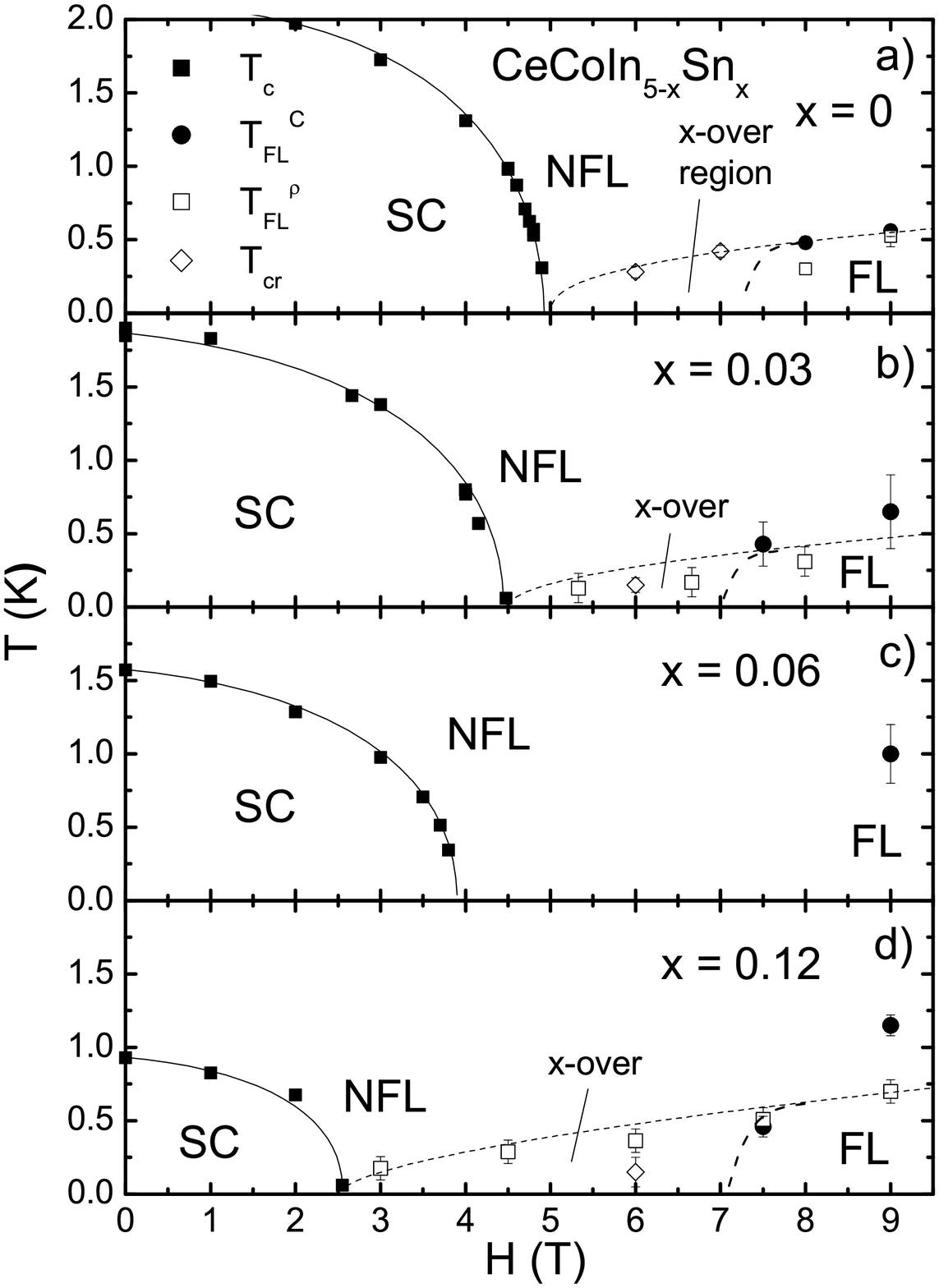}
\caption{Temperature-magnetic field ($T-H$) phase diagram of \CeCoSn{} for a) $x=0$ \cite{Bianchi03},   b) $x=0.03$, c) $x=0.06$, and d) $x=0.12$.  SC: superconducting; NFL: non-Fermi liquid; FL: Fermi liquid; x-over: crossover region.  The lines are guides to the eye. }
\label{phase}
\end{figure}

Figure \ref{phase} shows the magnetic field-temperature ($H-T$) phase diagrams for \CeCoSn{} for $x \leq 0.12$.   While the superconducting region is suppressed by Sn substitution in \CeCoIn,  NFL characteristics are observed in vicinity of the upper critical field for all $x$.  In particular, we do not observe the robust Fermi-liquid behavior near $H_{c2}$  that would be expected if the QCP was suppressed more rapidly than superconductivity.  Moreover,  no sign of an anomaly associated with magnetic order is found at $T=60$ mK in the magnetoresistance  (Fig. \ref{mr}a) [or $C(H,T)$]; such an anomaly is expected to occur if superconductivity was suppressed more rapidly than the critical point.  Thus, to within the width of the superconducting transition ($\sim0.5$ T), no long-range order is observed for $T \gtrsim 50$ mK and for $H\leq9$ T ($x \leq 0.12$).  The absence of FL behavior at $H_{c2}$ and long-range order provide further evidence for the occurrence of a QCP at $H_{c2}$ for all $x$. 
The $C(H,T)$ data reveal a crossover from NFL to FL behavior where the slope of the $-ln(T)$ dependence of $C_{el}/T$ decreases but does not saturate;  only for $H\geq7.5$ T does $C_{el}/T$ become constant, indicative of a FL ground state.  Electrical resistivity measurements reveal a similar picture with a (nearly) divergent $A$ coefficient near $H_{c2}$ for $x=0$ \cite{Paglione03}, 0.03, and 0.12, and  $T^2$ scattering over an extended temperature range in the FL region of the phase diagram.
%Fig 4

Having firmly established the existence of a quantum critical point at the upper critical field in \CeCoSn{} ($x \leq 0.12$), we conclude that the occurrence of quantum criticality and the destruction of superconductivity at $H_{c2} \approx H_{QCP}=5$ T in \CeCoIn{} is not mere coincidence, but is a signature of the underlying strongly correlated electron physics.  We  discuss two possible scenarios consistent with this novel type of quantum criticality.  An attractive scenario consistent with our data  is that of a superconducting quantum critical point. 
%In a conventional superconductor, the temperature range of superconducting fluctuations is quite small, of the order of a few mK; therefore, it is not  expected that such fluctuations would give rise to quantum criticality.  However, 
It has recently been shown that quantum criticality can arise in  a conventional BCS superconductor when pair breaking suppresses $T_c$ to zero temperature \cite{Ramazashvilli97}. In this case, the superconducting pair fluctuations are characterized by a dynamical critical exponent $z=2$, leading to singular corrections to the specific heat $\delta C/T \sim -ln(T/T_0)$ and electrical resistivity $\delta \rho \sim AT$ in two dimensions.   
Similar predictions for a superconducting quantum critical point developing from unconventional $d$-wave superconductivity are lacking at present, making direct comparison to experiment impossible.  However, both the fact that \CeCoIn{}  is a very clean superconductor \cite{Movshovich01}  and  the first-order nature of superconductivity near $H_{c2}$ \cite{Bianchi03b}, in which superconducting fluctuations are expected to be severely suppressed,   tend to preclude such a superconducting QCP  scenario.  

An alternative scenario is that superconductivity in \CeCoSn{} masks an unusual ordered state and an associated QCP.
Howell and Schofield \cite{Howell03} recently proposed a dissipative-fermion model at $T=0$ K in which a quantum critical point separates a Fermi-liquid metal from a non-Fermi liquid classical gas of particles with a finite zero-temperature entropy.  The quantum fluctuations of this unusual NFL state  are circumvented by the formation of superconductivity at finite temperature.  Such a scenario may, in fact, be realized in \CeCoSn; Sn substitution and/or magnetic field tune the quantum phase transition while superconductivity acts as a veil  that is parasitic to the abundant low-energy quantum fluctuations.  Once the underlying phase is destroyed at $H_{QCP}=H_{c2}$, the protective envelope of superconductivity is no longer necessary and the system exhibits critical behavior in the vicinity of the QCP, leading to the phase diagram shown in Fig. \ref{phase}.  While it is not clear whether fluctuations of the underlying quantum phase transition mediate the superconductivity encompassing it, we conjecture that quantum fluctuations in vicinity to a hidden antiferromagnetic quantum critical point provide a natural explanation for $d$-wave superconductivity in \CeCoSn.  This picture is in agreement with recent   thermal and charge transport  measurements in magnetic fields on \CeCoIn{}  suggesting that the critical fluctuations are magnetic in nature \cite{Paglione04}.

There is evidence for a similar superconducting ``veil" in another heavy fermion compound UBe$_{13}$ \cite{Steglich98}.   In this material, a divergent Sommerfeld coefficient at $H_{c2}=12$ T and a decrease of $A$ away from $H_{c2}$  are observed  \cite{Steglich98}, identical to what is found in \CeCoSn.  This picture is qualitatively different from other heavy fermion systems (e.g., CePd$_2$Si$_2$)  where antiferromagnetic order is suppressed by the application of pressure and the QCP lies well within the superconducting dome \cite{Mathur98}. It is an open question whether the two types of phase diagrams comprise two separate, unrelated situations, or if, in fact, they are manifestations of the same underlying physics that is governed by the relative strengths of the two phenomena at ambient conditions.  Further measurements are necessary to elucidate these issues.

In summary, $C(H,T)$ and $\rho(H,T)$  measurements performed on \CeCoSn{}   are consistent with a QCP located at $H_{c2}$  for all $x\leq 0.12$. This novel behavior  in \CeCoSn{} is  most likely associated with an underlying (antiferromagnetic) phase transition  masked by unconventional superconductivity  and probably cannot be accounted for within a superconducting QCP scenario.     
We hope this work stimulates further experimental and theoretical investigations of quantum criticality associated with  unconventional superconductivity.

Work at Los Alamos was performed under the auspices of the U.S. DOE.  We thank A. Millis, R. Ramazashvilli, A. Schofield, G. Sparn, and I. Vekhter for helpful discussions. F. R. thanks the Reines Fellowship (DOE-LANL) for support.

% \bibliography{bibliv2}
% \bibliographystyle{apsrev}

\end{document}